\documentclass[letter]{aa}

\usepackage{graphicx}
\usepackage{txfonts}
\usepackage{amsmath}
\usepackage{booktabs}
\usepackage{multirow}
\usepackage{microtype}
\usepackage{float}

\newcommand{\mbh}{M_{\rm BH}}
\newcommand{\ledd}{\lambda_{\rm Edd}}

\newcommand{\sfinf}{{\rm SF}_{\infty}}

\begin{document}

\title{Rest-frame Ultraviolet Variability of Quasars in the Cosmic Himalayas Overdensity}

\author{
Jiaqi Lin\inst{1,2,3}\email{linjq63@mail2.sysu.edu.cn}
\and
Zhenya Zheng\inst{2}\corrauth{zhengzy@shao.ac.cn}
\and
Bin Ma\inst{1,3}\email{mabin3@mail.sysu.edu.cn}
\and
Yongming Liang\inst{4,5}\email{ymliang@icrr.u-tokyo.ac.jp}
\and
Lin Long\inst{2,6}\email{longlin@shao.ac.cn}
}

\institute{
School of Physics and Astronomy, Sun Yat-sen University,
Zhuhai 519082, People's Republic of China
\and
Shanghai Astronomical Observatory, Chinese Academy of Sciences,
80 Nandan Road, Shanghai 200030, People's Republic of China
\and
CSST Science Center for the Guangdong-Hong Kong-Macau Greater Bay Area,
Zhuhai 519082, People's Republic of China
\and
Institute for Cosmic Ray Research, The University of Tokyo, 5-1-5 Kashiwanoha, Kashiwa, Chiba 277-8582, Japan
\and
National Astronomical Observatory of Japan, 2-21-1 Osawa, Mitaka, Tokyo 181-8588, Japan
\and
School of Astronomy and Space Sciences, University of Chinese Academy of Sciences, No. 19A Yuquan Road, Beijing 100049, People's Republic of China
}

\date{Received September 30, 20XX}

\abstract
{Some high-redshift overdensities show an enhanced incidence of active galactic nuclei (AGNs), suggesting that large-scale environment may influence supermassive black hole fuelling and triggering. Whether environment also affects AGN variability remains unclear.}
{Using the Cosmic Himalayas (CH), an extreme concentration of 11 type 1 quasars at $z=2.16$--$2.20$, we investigate possible offsets in rest-frame ultraviolet variability relative to quasars with comparable redshifts, black hole masses, and Eddington ratios.}
{We construct a parent sample of 1,138 Stripe 82 quasars at $2.11\leq z\leq2.25$ and combine photometry from the Sloan Digital Sky Survey (SDSS), the Panoramic Survey Telescope and Rapid Response System 1 (PS1), and the Zwicky Transient Facility (ZTF) to build observer-frame $g$- and $r$-band light curves spanning up to 27~yr. We compare damped random-walk (DRW) parameters and ensemble structure functions (SFs) with control pseudo-samples matched in black hole mass and Eddington ratio, and use injection--recovery simulations to forecast the detectability of a CH-like offset.}
{The CH quasars show a tendency toward shorter DRW damping timescales (median offsets of $-0.32$ and $-0.13$ dex in $g$ and $r$) and slightly higher asymptotic amplitudes. None of the four DRW comparisons is statistically significant, including the largest offset in the $g$-band damping timescale ($p=0.16$). Ensemble SFs generally exceed matched-control medians, although 18 of 19 lag bins per band remain within the 95\% pointwise envelopes. Assuming a true uniform $-0.32$ dex $g$-band timescale offset, injection--recovery simulations predict an 80\% probability of nominal $3\sigma$ detection with approximately 60 quasars of comparable variability scatter and data quality.}
{If confirmed, the tentative variability offsets may suggest that large-scale environment influences accretion-flow variability beyond its potential role in quasar triggering. Larger samples with higher-precision light curves, such as those expected from the Legacy Survey of Space and Time (LSST), are required to test this possibility.}


\maketitle
\nolinenumbers  

\section{Introduction}\label{sec:intro}

At $z\simeq2$--3, protoclusters and other large-scale overdensities provide important laboratories for studying environmental effects on galaxy evolution and supermassive black hole (SMBH) growth. By influencing gas supply and galaxy interactions, these environments may affect the fuelling and triggering of active galactic nuclei (AGNs), and several high-redshift overdensities show enhanced AGN fractions or X-ray luminosity functions \citep{Lehmer2009,Tozzi2022,Shah2025}, although such enhancement is not universal \citep{Macuga2019}. These studies mainly probe the incidence of AGN activity. On smaller scales, optical variability has been found to anticorrelate with near-infrared covering factor, suggesting a connection to the close dusty environment \citep{Wu2024}. Whether quasar variability also depends on environment on much larger scales remains unclear.

The Cosmic Himalayas (CH) provides an extreme setting in which to address this question. It is a concentration of 11 type~1 quasars at $z=2.16$--$2.20$ within a volume of approximately $(40~{\rm cMpc})^3$ \citep{Liang2025}. Across the ${\sim}10,000~{\rm deg}^{2}$ SDSS footprint examined by \citet{Liang2025}, the CH represents the highest quasar-density peak, with a quasar overdensity of $\delta_{\rm QSO}\approx30$ and a reported significance of $16.9\sigma$. It lies near the midpoint of a ${\sim}100$~cMpc Ly$\alpha$ emitter (LAE) filament connecting two overdensity nodes, while the quasar-density peak is offset from both nodes and coincides with neither an intergalactic medium (IGM) \ion{H}{i} absorption nor transmission peak. The CH therefore represents an unusually concentrated quasar population embedded in a complex large-scale environment. We use this extreme quasar concentration as a case study to investigate possible offsets in rest-frame ultraviolet (UV) variability relative to matched quasars.

Quasars commonly exhibit stochastic UV/optical variability, which is
often described by a damped random walk (DRW), characterised by a damping timescale, $\tau_{\rm DRW}$, and an asymptotic variability amplitude, $\sfinf$ \citep{Kelly2009,MacLeod2010,Suberlak2021ApJ}. The structure function (SF) provides a complementary description of variability as a function of time lag without assuming a DRW form. Previous studies have found that quasar variability properties are associated with black hole mass, Eddington ratio, and rest-frame wavelength \citep{Kelly2009,MacLeod2010,Burke2021}. Their inferred values are also sensitive to the duration and sampling of the light curves, particularly for long characteristic timescales \citep{Stone2022,Paolillo2025}. These known correlations and observational effects need to be considered when assessing whether the CH quasars have unusual variability properties.

For this study, we combine multi-epoch photometry from the Sloan Digital Sky Survey \citep[SDSS;][]{York-2000AJSDSS}, the Panoramic Survey Telescope and Rapid Response System 1 Survey \citep[PS1;][]{Chambers-2016arXivpanstarr}, and the Zwicky Transient Facility \citep[ZTF;][]{Bellm-2019PASPZTF,Masci-2019PASPZTF} to construct observer-frame $g$- and $r$-band light curves spanning up to 27~yr. At $z\simeq2.2$, these bands probe the rest-frame UV wavelengths at approximately $1500$--$2000\,\text{\AA}$. We define a Stripe 82 (S82) parent sample over $2.11 \leq z \leq 2.25$ and select control quasars matched in black hole mass and Eddington ratio. We then compare the CH quasars with the matched controls using both the DRW parameters and the ensemble SF. We use these comparisons to investigate whether the CH quasars exhibit a population-level difference in variability relative to the matched controls. Throughout this study, we adopt $H_{0}$=70~$\rm{km\ s^{-1}\ Mpc^{-1}}$, $\rm{\Omega}_{m}=0.3$ and $\rm{\Omega_{\Lambda}=0.7}$.

\section{Data and sample}\label{sec:data}

\subsection{CH quasars and the S82 parent sample}\label{sec:sample}

The CH sample comprises 11 type 1 quasars at $2.16 \leq z \leq 2.20$ identified by \citet{Liang2025}, all of which lie within the S82 equatorial field. We construct the parent comparison sample from the SDSS Data Release 16 Quasar Catalogue \citep[DR16Q;][]{Lyke-2020} within the same field. S82 is a repeatedly observed equatorial field covering approximately $275~\mathrm{deg}^{2}$, spanning $-50^{\circ} \lesssim \alpha_{\mathrm{J2000}} \lesssim 60^{\circ}$ and $-1.25^{\circ} \lesssim \delta_{\mathrm{J2000}} \lesssim 1.25^{\circ}$ \citep{Annis-2014ApJ}.

All 11 CH quasars are removed before the S82 parent sample is defined. We restrict the remaining S82 quasars to $2.11\leq z\leq2.25$, so that the observer-frame $g$ and $r$ bands probe nearly the same rest-frame UV wavelength ranges as in the CH quasars. Every retained object has usable $g$- and $r$-band light-curve data from SDSS, PS1, and ZTF. Because ZTF is shallower than SDSS and PS1, it sets the effective photometric limit of the combined light curves. We therefore first require median ZTF magnitudes of $g_{\mathrm{med}}<21.5$ and $r_{\mathrm{med}}<21.5$~mag (AB system) to ensure sufficient signal-to-noise in the ZTF photometry. Within this magnitude-limited sample, we further require at least 20 nightly SDSS measurements and 20 nightly ZTF measurements in each band. These criteria yield a final S82 parent sample of $1\,138$ quasars.

\begin{figure}[t]
\centering
\includegraphics[width=\columnwidth]{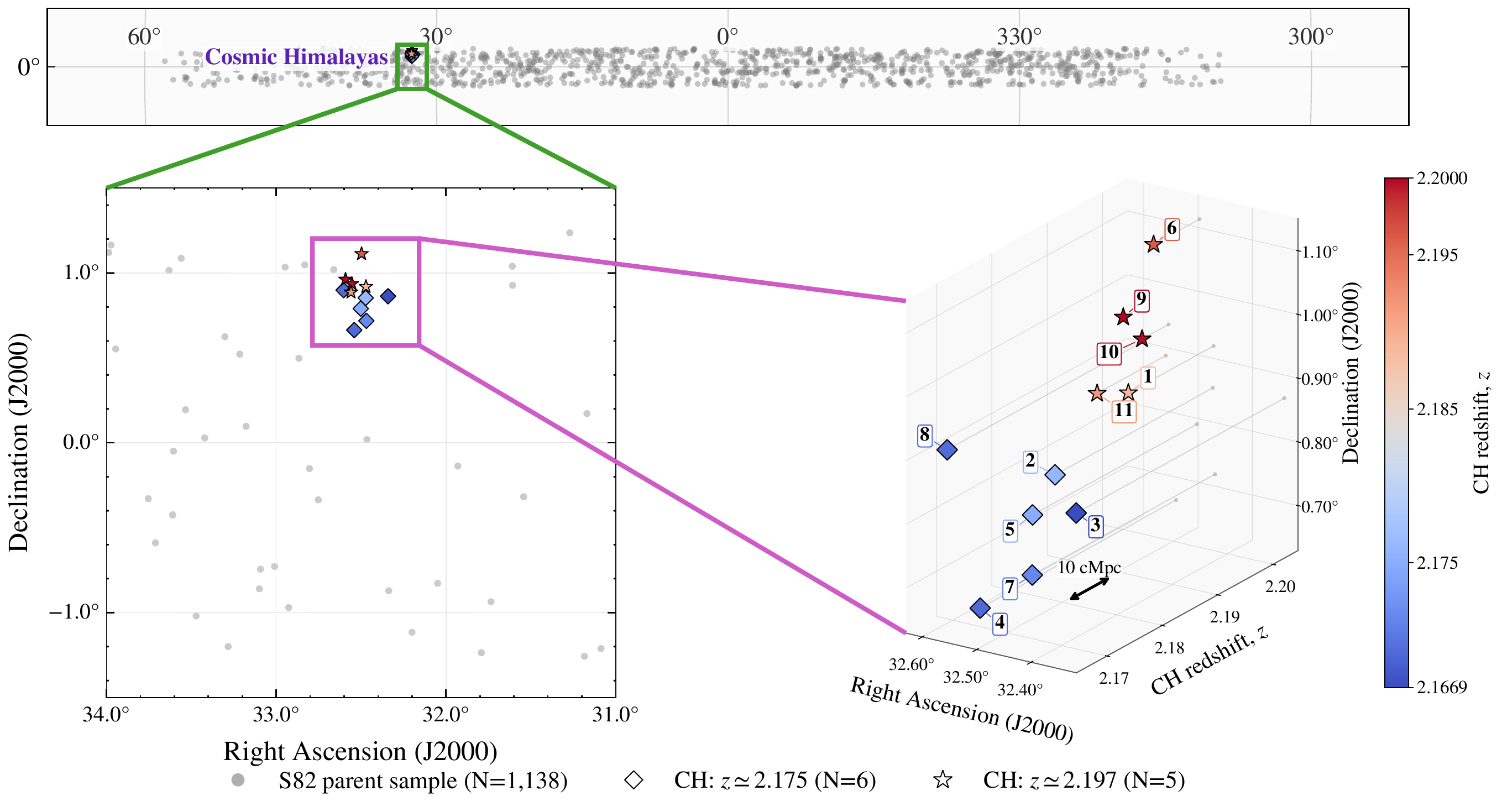}
\caption{Spatial distribution of the CH and S82 samples. Grey points represent the S82 parent sample ($N=1,138$). The 11 CH quasars are marked by numbered diamonds ($z\simeq2.175$, $N=6$) and stars ($z\simeq2.197$, $N=5$), with colors indicating their redshifts.}
\label{fig:parent}
\end{figure}

We adopt the CH quasar redshifts from \citet{Liang2025}, except for CH-Q02. For this source, we adopt $z=2.176$ based on Keck/MOSFIRE H$\alpha$ spectroscopy (Liang et al., in prep.). For the S82 parent sample, redshifts are taken from the DR16Q quasar-property catalogue of \citet{Wu-2022ApJS}, which also provides the single-epoch virial black hole masses, bolometric luminosities, and Eddington ratios for both samples. The adopted black hole masses are based on C IV and may be affected by outflows, introducing uncertainties into the sample matching. The spatial distributions of the CH quasars and S82 parent sample within the S82 footprint are shown in Fig.~\ref{fig:parent}. The 11 CH quasars comprise two redshift subgroups at $z\simeq2.175$ ($N=6$) and $z\simeq2.197$ ($N=5$).


\subsection{Light-curve construction}\label{sec:lc}

We combine photometry from SDSS (1998--2007), PS1 (2010--2014), and ZTF (2018--2025) to construct observer-frame $g$- and $r$-band light curves for quasars in the CH and S82 parent samples, providing baselines of up to approximately 27 yr in the observer frame. For each survey, we remove low-quality measurements and photometric outliers, combine repeated observations within the same night, and transform the PS1 and ZTF photometry onto the SDSS photometric system. The S82 quasars use nightly ZTF catalogue measurements. Individual ZTF exposures are generally too shallow for the fainter CH quasars, so we derive their ZTF photometry from adaptively time-binned, weighted image coadds following \citet{Lin2025ApJS}. The quality cuts, nightly combination, cross-survey photometric homogenisation, and image-coaddition procedure are described in Appendix~\ref{app:lc}.

\section{Analyses and results}\label{sec:results}

\subsection{The CH quasars in the parent variability space}
\label{sec:parent_variability}

We characterise the stochastic variability of each quasar with a DRW, a stationary Gaussian process commonly used to describe quasar optical variability \citep{Kelly2009,MacLeod2010}. We model the $g$- and $r$-band light curves independently with the DRW fitting implemented in the DRW-based outlier-rejection module of \texttt{pyPetal} \citep{Stone2024}. The module uses \texttt{celerite} for Gaussian-process modelling \citep{ForemanMackey2017,ForemanMackey2018} and \texttt{emcee} for posterior sampling \citep{ForemanMackey2013}. For each DRW parameter, we adopt the posterior median as the parameter estimate, with the 16th and 84th percentiles defining the lower and upper uncertainties. We use the rest-frame damping timescale, $\tau_{\rm rest}$, and the asymptotic variability amplitude, $\sfinf$, to characterise the variability timescale and amplitude, respectively. Their definitions and the relation between the DRW covariance and $\sfinf$ are given in Appendix~\ref{app:drw}.

As shown in Fig.~\ref{fig:parentdrw}, the CH quasars broadly overlap the S82 parent-sample distributions of $\tau_{\rm rest}$ and $\sfinf$, including their relations with $M_{\rm BH}$ and $\lambda_{\rm Edd}$ in both bands. The medians of the two redshift subgroups also lie within these broad distributions. Since the parent sample is not matched in these properties, we use the matched-control comparison below to quantify the magnitude and statistical significance of group-level offsets.

\begin{figure}[t]
\centering
\includegraphics[width=1.0\columnwidth]{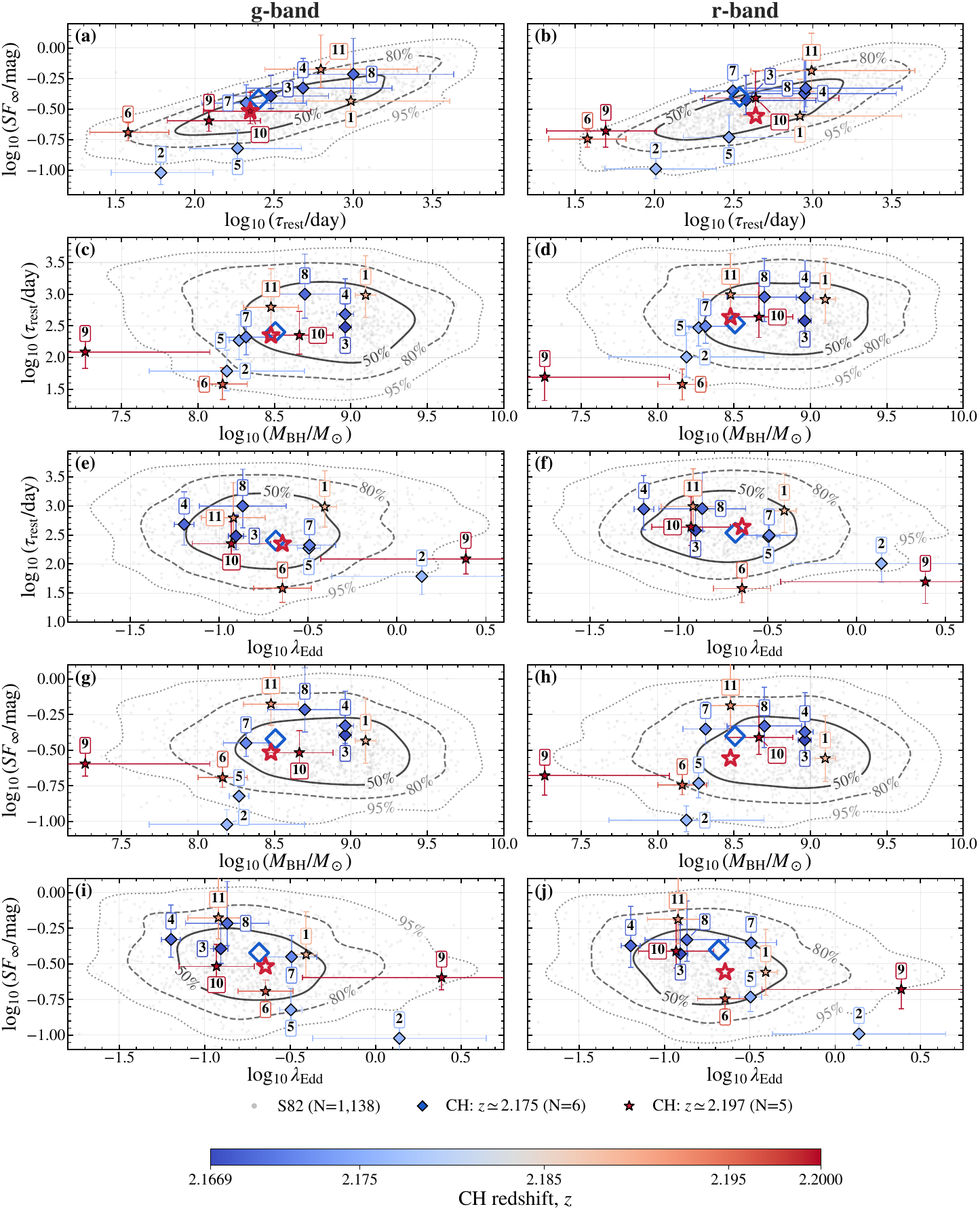}
\caption{Location of the CH quasars in the variability parameter space of the S82 parent sample. The left and right columns correspond to the $g$ and $r$ bands, respectively. Panels (a,b) show the $\tau_{\rm rest}$--$\sfinf$ distributions. Panels (c--f) and (g--j) show $\tau_{\rm rest}$ and $\sfinf$ versus $M_{\rm BH}$ and $\lambda_{\rm Edd}$, respectively. Error bars on the DRW parameters indicate their 16th--84th percentile posterior intervals. In panels (c--j), horizontal error bars indicate the catalogue uncertainties in $M_{\rm BH}$ and $\lambda_{\rm Edd}$. Open symbols indicate the medians of the two CH redshift subgroups.}
\label{fig:parentdrw}
\end{figure}

\subsection{Matched-control comparison of the DRW parameters}
\label{sec:drwresults}

Quasar variability properties correlate with black hole mass and Eddington ratio \citep{Wilhite2008,Kelly2009,MacLeod2010}. We therefore standardise $\log_{10}(\mbh/M_{\odot})$ and $\log_{10}\ledd$ using the S82 parent sample and define the 50 nearest neighbours of each CH quasar as its control pool. We then generate $10\,000$ control pseudo-samples by drawing one quasar from each of the 11 control pools, with no duplicate controls within a pseudo-sample. For each DRW parameter, we calculate the median of the 11 controls in each pseudo-sample. The resulting $10\,000$ medians form the matched-control reference distribution. The full matching and group-level comparison procedure is described in Appendix~\ref{app:matching_method}, and the control-pool balance diagnostics are presented in Appendix~\ref{app:matching}.

As shown in Fig.~\ref{fig:nulls}, the CH quasars show a weak tendency toward shorter damping timescales and slightly larger asymptotic variability amplitudes relative to the matched-control reference medians, with the largest offset in the $g$-band damping timescale ($-0.32$ dex). However, none of the four comparisons is statistically significant (empirical two-sided $p=0.16$--$0.88$).

Because the CH ZTF photometry is measured from image coadds whereas the S82 controls use nightly catalogue measurements, the two samples have different temporal sampling. To assess whether this difference affects the DRW comparison, we rebin the ZTF light curves of the S82 controls into 35, 90, and 150~d bins and repeat the analysis. The reference distributions change only modestly, and all empirical two-sided $p$ values remain above 0.05. The qualitative DRW comparison is unchanged under the tested ZTF rebinning schemes. Further details are given in Appendix~\ref{app:binning}.

\begin{figure}[t]
\centering
\includegraphics[width=\columnwidth]{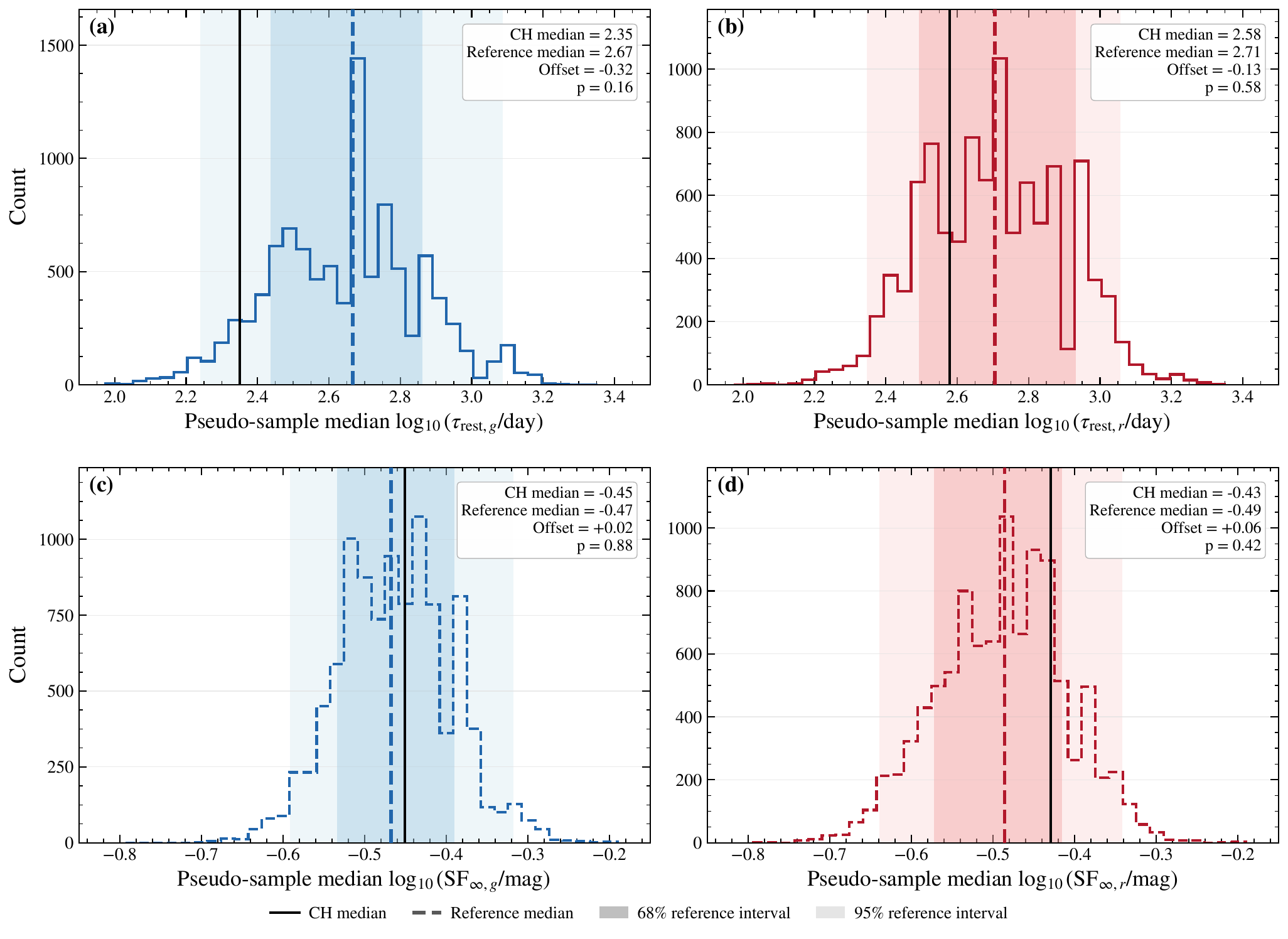}
\caption{Matched-control reference distributions of the median DRW parameters. Panels~(a) and (b) show $\log_{10}(\tau_{\rm rest}/{\rm d})$ in the $g$ and $r$ bands, and panels~(c) and (d) show $\log_{10}(\sfinf/{\rm mag})$. Solid black lines mark the CH medians, dashed coloured lines mark the reference medians, and the dark and light shaded regions indicate the central 68\% and 95\% reference intervals.}
\label{fig:nulls}
\end{figure}

\subsection{Ensemble SF}\label{sec:sf}

To complement the DRW-based analysis, we compare the ensemble SFs of the CH quasars with those of the matched controls across a range of rest-frame time lags. In each rest-frame lag bin, the CH ensemble SF is defined as the median of the valid source-level SF values, with its sampling uncertainty estimated by bootstrap resampling. We apply the same median estimator to the $10\,000$ matched-control pseudo-samples to construct the matched-control reference distribution and its 68\% and 95\% pointwise envelopes. The SF estimator and resampling procedure are described in Appendix~\ref{app:sf}.

As shown in Fig.~\ref{fig:sf}, the CH ensemble SFs show a tendency to lie above the matched-control medians across multiple rest-frame lag bins. In each band, the only point outside the 95\% pointwise envelope corresponds to the rest-frame lag bin spanning 30--95~d.

\begin{figure}[t]
\centering
\includegraphics[width=\columnwidth]{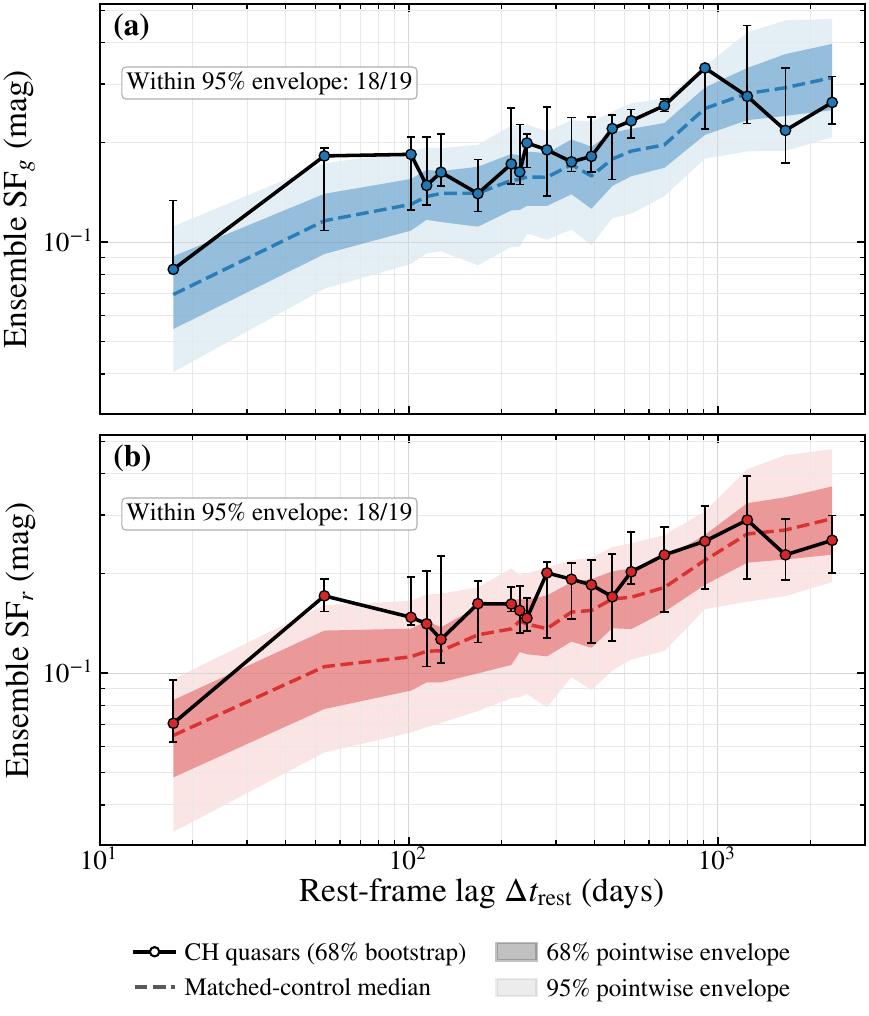}
\caption{Ensemble SFs in the (a) $g$ and (b) $r$ bands. Points and solid lines show the CH ensemble medians, and the error bars show the 16th--84th percentile intervals from source-level bootstrap resampling. Dashed lines show the medians of the matched-control pseudo-sample distributions, while the dark and light shaded regions show the 68\% and 95\% pointwise envelopes, respectively. The annotations give the number of lag bins lying within the 95\% envelopes.}
\label{fig:sf}
\end{figure}

\section{Discussion and conclusions}\label{sec:discussion}

To investigate whether the CH quasars show unusual rest-frame UV variability, we compare the 11 CH quasars with matched S82 quasars using DRW parameters and ensemble SFs. The CH sample shows a tendency toward shorter DRW damping timescales, with the largest offset occurring in the $g$ band ($-0.32$ dex), while the asymptotic amplitudes are only slightly higher than the matched-control reference values. None of the four DRW comparisons is statistically significant. The ensemble SFs also tend to lie above the matched-control medians across multiple rest-frame lag bins, although 18 of the 19 bins in each band remain within the corresponding 95\% pointwise envelopes. 


Our work extends studies of AGNs in high-redshift overdensities to the time domain, providing a complementary view through long-term variability. Previous studies suggest that overdense environments may influence AGN fuelling or triggering even when the bulk properties of AGNs remain similar to those in the field. In SSA22, \citet{Monson2023} found that the SMBH growth rate, black hole mass, star-formation rate, and stellar mass distributions of protocluster AGNs are broadly consistent with those of field AGNs, while AGN triggering may be more frequent in the protocluster. Environmental influences on AGN fuelling and triggering may arise through differences in gas supply and galaxy interactions. In three $z\sim2$ protoclusters, \citet{Tadaki2019PASJ} found that galaxies with stellar masses of $10^{10.5}$--$10^{11},M_\odot$ have higher molecular-gas fractions and longer depletion times than expected from field scaling relations, consistent with environment-dependent gas accretion. In SSA22 at $z=3.1$, \citet{Hine2016MNRAS} measured a merger fraction of $48\pm10\%$ for protocluster Lyman-break galaxies, compared with $30\pm6\%$ in the field. Although this excess is only marginal, it suggests that galaxy interactions may be more frequent in overdense environments. These differences could modify nuclear fuelling conditions. On smaller scales, \citet{Wu2024} found a connection between optical variability and the circumnuclear dusty environment, further suggesting that quasar variability may be sensitive to conditions surrounding the central SMBH.

In this context, our results have two possible physical implications for the CH. If the tentative trend toward shorter DRW damping timescales is confirmed, it may suggest that large-scale environment is linked to accretion-flow variability. Such a connection could arise from environmental differences in gas supply or galaxy interactions that may modify nuclear fuelling conditions. Conversely, if CH quasars continue to show variability consistent with matched field quasars, the extreme quasar overdensity could partly reflect a higher incidence of quasar activity. This may arise from more frequent triggering or a higher active duty cycle, without requiring systematically different variability among active quasars. This scenario is qualitatively consistent with the interpretation proposed by \citet{Monson2023} for AGNs in the SSA22 protocluster.

\begin{figure}[t]
\centering
\includegraphics[width=\columnwidth]{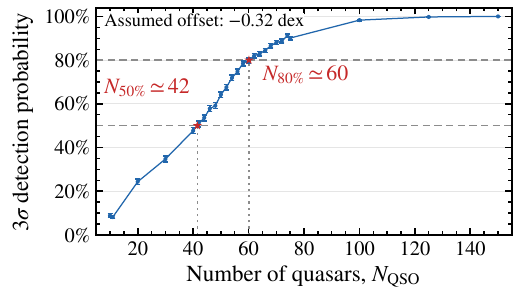}
\caption{Detection probability versus quasar sample size, assuming a true CH-like offset of $-0.32$ dex in the $g$-band rest-frame DRW damping timescale. The blue curve shows the fraction of 5000 mock samples per sample size, drawn from matched S82 controls, yielding a two-sided empirical $p<0.0027$ (nominal $3\sigma$). Error bars denote 95\% Wilson Monte Carlo intervals. Red stars mark the sample sizes corresponding to 50\% and 80\% detection probabilities, approximately 42 and 60 quasars, respectively.}
\label{fig:power_vs_N_QSO}
\end{figure}

To assess the prospects for testing the tentative offset, we performed injection--recovery simulations using the matched S82 controls, assuming a uniform offset of $-0.32$ dex in the $g$-band rest-frame damping timescale. As shown in Fig.~\ref{fig:power_vs_N_QSO}, approximately 42 and 60 quasars would provide 50\% and 80\% probabilities of achieving a nominal $3\sigma$ detection, respectively. These estimates assume that the observed offset represents the true population difference and that future samples have comparable variability scatter and measurement quality.

Larger samples with higher-precision light curves will therefore be needed to determine whether such a connection exists. The Legacy Survey of Space and Time (LSST) at the Vera C. Rubin Observatory will deliver precisely such a dataset, providing high-quality multiband light curves for millions of sources over a 10-year survey with an average cadence of 3 days in six photometric bands \citep{Ivezic2019}.

\begin{acknowledgements}
Z.Y.Z. acknowledges the support of the Shanghai Leading
Talent Program of Eastern Talent Plan (LJ2025051) and the
China-Chile Joint Research Fund 178GJHZ2025067MI. We also acknowledge the science research grants from the China Manned Space Project, especially, NO.  CMS-CSST-2025-A18. YL acknowledges the support from the JSPS KAKENHI grants No. 24K17084 and 25H00663.\\

\textit{Author contributions.}
J.Q.L led the project and performed the data analysis. Z.Y.Z contributed to the original concept and helped define the overall scientific framework. All authors discussed the results and provided comments on the manuscript.
\end{acknowledgements}

\bibliographystyle{aa}
\bibliography{references}

\appendix
\nolinenumbers

\section{Construction of the multi-survey light curves}\label{app:lc}

\subsection{Survey data processing and photometric homogenisation}\label{app:multi-survey}

Multi-epoch $g$- and $r$-band light curves are constructed from SDSS, PS1, and ZTF photometry. A matching radius of $1\arcsec$ is adopted for all three surveys, with survey-specific quality cuts applied before transforming the measurements to a common photometric system. For SDSS, we use single-epoch detections from the \texttt{PhotoObj} table in the \texttt{Stripe82} context of SDSS CasJobs\footnote{\url{https://skyserver.sdss.org/casjobs/}}, retaining only detections with the SDSS \texttt{clean} flag set (\texttt{clean}=1). For PS1, we identify counterparts in the \texttt{MeanObject} table and retrieve the associated single-epoch PSF photometry from the \texttt{Detection} table through MAST CasJobs\footnote{\url{https://mastweb.stsci.edu/mcasjobs/}}. The PSF fluxes are converted to AB magnitudes, with measurements required to satisfy \texttt{psfQfPerfect}>0.95, where \texttt{psfQfPerfect} is the PSF-weighted fraction of fully unmasked pixels. For the S82 quasars, we obtain ZTF DR24 catalogue light curves through IRSA\footnote{\url{https://irsa.ipac.caltech.edu/data/ZTF/docs/releases/dr24/}}, require \texttt{catflags}=0, and retain detections from the field that provides the largest number of epochs for each source. Across all three surveys, we reject measurements with magnitude uncertainties greater than 0.25~mag.

We further clean the SDSS light curves of both samples and the ZTF light curves of the S82 quasars separately in each band. Gaps longer than 120~d define separate observing seasons. Within each season, we remove detections that deviate from the median by more than three times the standard deviation of the seasonal light curve. Measurements obtained during the same night are then combined using inverse variance weights. The ZTF measurements of the CH quasars are derived from image coadds and are processed separately as described in Appendix~\ref{app:ztf-coadd}.

We use spectral energy distribution (SED) modelling and synthetic photometry to derive an individual correction for each quasar and transform its PS1 and ZTF measurements to the SDSS photometric system. Specifically, we fit the deep coadded S82 $ugriz$ photometry from \citet{Jiang2014} with the parametric type 1 quasar SED model implemented in \texttt{qsogen}\footnote{\url{https://github.com/MJTemple/qsogen}} \citep{Temple2021}. If the fit does not converge, we adopt the model spectrum provided by the SDSS pipeline for that source. We then calculate synthetic $g$ and $r$ magnitudes by integrating the adopted spectrum through the response curves of SDSS, PS1, and ZTF. In each band, the differences between the synthetic PS1 or ZTF magnitude and the corresponding synthetic SDSS magnitude define the required corrections.

\subsection{The ZTF image coaddition for the CH quasars}\label{app:ztf-coadd}

\begin{figure*}[p]
\centering
\includegraphics[height=0.88\textheight]{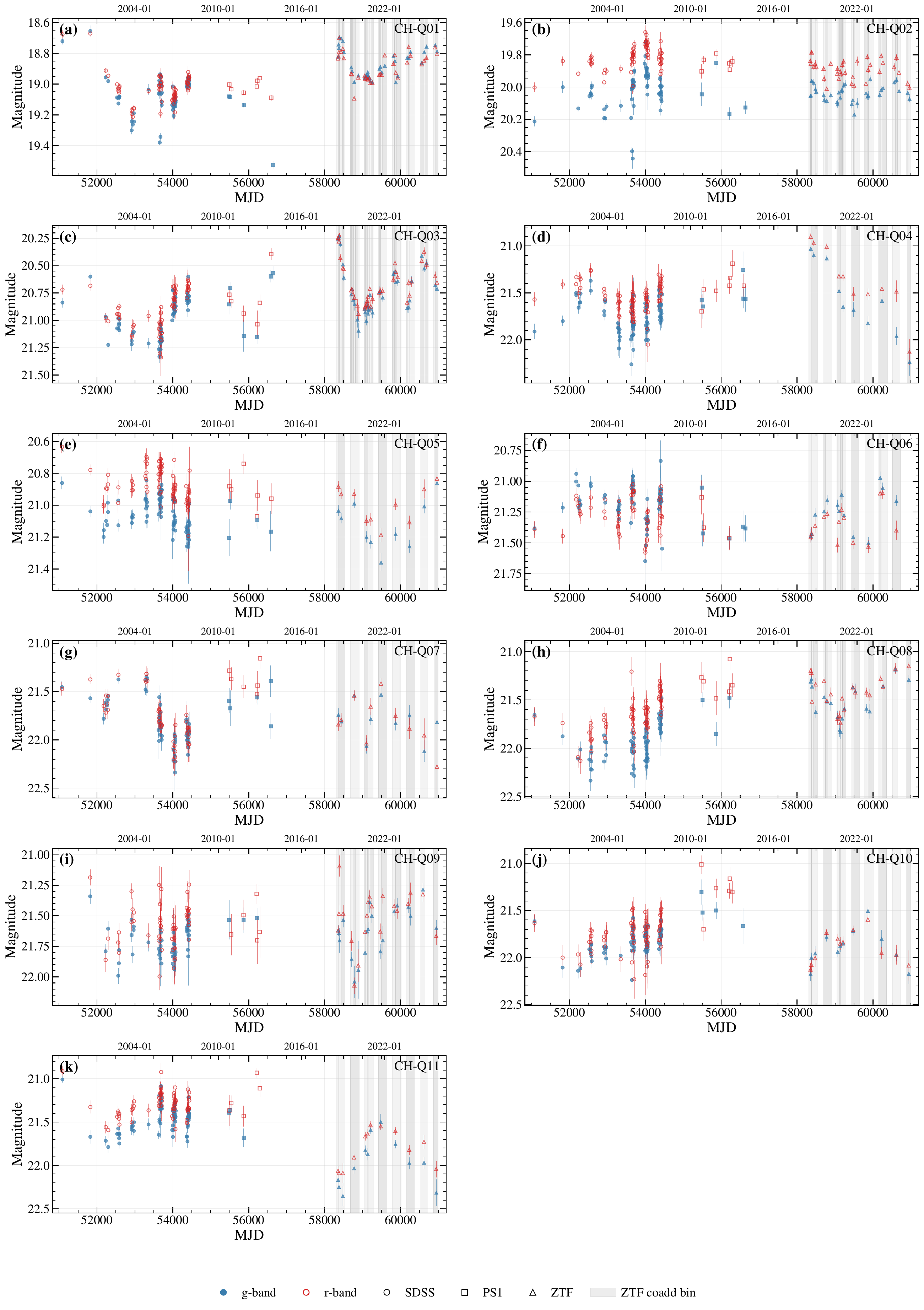}
\caption{Multi-survey $g-$ and $r-$band light curves of the 11 CH quasars. Panels~(a)--(k) show the individual sources. Blue and red symbols represent the $g$ and $r$ bands, respectively. Circles, squares, and triangles denote measurements from SDSS, PS1, and the ZTF image coadds, respectively. The grey shaded intervals show the time spans of the exposures contributing to each ZTF coadd, with each triangle plotted at the median observation time of the corresponding exposures. The PS1 and ZTF measurements are transformed to the SDSS photometric system.}
\label{fig:lightcurves}
\end{figure*}

Most CH quasars are too faint for reliable photometry from individual 30~s ZTF exposures. We therefore construct image coadds in adaptive time bins for the $g$ and $r$ bands following \citet{Lin2025ApJS}. After quality filtering, the exposures within each bin are astrometrically aligned, scaled to a common photometric level, and combined using weights that account for transparency, seeing, sky noise, and the fraction of masked pixels. The minimum number of exposures per coadd ranges from 15 to 40, depending on the target and band, to balance photometric depth and temporal resolution. We perform PSF photometry on each coadded image and determine its photometric zero point using reference sources matched to the PS1 DR1 \texttt{MeanObject} catalogue. Each coadded measurement is assigned the median observation time of its contributing exposures. Finally, we transform the ZTF magnitudes to the SDSS photometric system using the correction derived for each source. The resulting light curves are shown in Fig.~\ref{fig:lightcurves}.

\section{Variability measurements}\label{app:variability}

\subsection{DRW modelling}\label{app:drw}

Following the
parameterisation of \citet{MacLeod2010}, the DRW process is characterised
by an exponential covariance function,
\begin{equation}
\kappa(\Delta t)
=
\sigma_{\rm DRW}^{2}
\exp\left(-\frac{|\Delta t|}{\tau_{\rm DRW}}\right),
\end{equation}
where $\Delta t$ is the time lag between two observations, $\tau_{\rm DRW}$ is the characteristic damping timescale, and $\sigma_{\rm DRW}$ is the stationary rms amplitude of the process. The corresponding first-order SF is
\begin{equation}
{\rm SF}^{2}(\Delta t)
=
\left\langle
\left[m(t+\Delta t)-m(t)\right]^{2}
\right\rangle.
\end{equation}
For a stationary process, the SF is related to the covariance function through
\begin{equation}
{\rm SF}^{2}(\Delta t)
=
2\left[\kappa(0)-\kappa(\Delta t)\right].
\end{equation}
Substituting the DRW covariance function gives
\begin{equation}
{\rm SF}(\Delta t)
=
\sqrt{2}\,\sigma_{\rm DRW}
\left[
1-\exp\left(-\frac{|\Delta t|}{\tau_{\rm DRW}}\right)
\right]^{1/2}.
\end{equation}
At time lags much longer than the damping timescale, $\Delta t\gg\tau_{\rm DRW}$, the SF approaches its asymptotic amplitude,
\begin{equation}
\sfinf
\equiv
\lim_{\Delta t\rightarrow\infty}{\rm SF}(\Delta t)
=
\sqrt{2}\,\sigma_{\rm DRW}.
\end{equation}

For each light curve, we adopt the posterior median as the parameter estimate of each DRW parameter. The observer-frame damping timescale is converted to the quasar rest frame as
\begin{equation}
\tau_{\rm rest}
=
\frac{\tau_{\rm DRW}}{1+z}.
\end{equation}
The resulting $\tau_{\rm rest}$ and $\sfinf$ values are used throughout the population-level comparisons.

\subsection{Ensemble SF calculation}\label{app:sf}

For source $s$ and rest-frame lag bin $k$, we calculate
\begin{equation}
{\rm SF}_{s}^{2}(\Delta t_k)=
\frac{1}{N_{{\rm pair},s,k}}
\sum_{\substack{i<j\\ \Delta t_{ij,{\rm rest}}\in k}}
\left[(m_i-m_j)^2-\sigma_i^2-\sigma_j^2\right],
\end{equation}
where $m_i$ and $m_j$ are the magnitudes, $\sigma_i$ and $\sigma_j$ are their photometric uncertainties, and $N_{{\rm pair},s,k}$ is the number of measurement pairs in the bin. The observer-frame time lags are converted to the rest frame by dividing by $(1+z)$. A source-level SF is retained when $N_{{\rm pair},s,k}\geq6$ and ${\rm SF}_{s}^{2}>0$, with ${\rm SF}_{s}=\sqrt{{\rm SF}_{s}^{2}}$. Source-level measurements that do not satisfy these criteria are excluded from the ensemble calculation.

For each band and lag bin, the CH ensemble SF is the median of the valid source-level SF values, provided that at least three CH quasars have valid measurements. We estimate the sampling uncertainty of this median by non-parametric bootstrap resampling of the valid source-level SF values. If $n$ CH quasars have valid SF values in a given bin, we draw $n$ values with replacement and calculate their median. This procedure is repeated $10\,000$ times to obtain the bootstrap distribution of the ensemble median, whose 16th and 84th percentiles define the plotted 68\% interval.

For the matched controls, each pseudo-sample contains 11 controls selected as described in Appendix~\ref{app:matching_method}. For each pseudo-sample, the ensemble SF is the median of the valid control SF values in each band and lag bin, provided that at least three are available. The $10\,000$ pseudo-samples therefore define the matched-control distribution in each band and lag bin. The medians of these distributions form the reference curve, while their 16th--84th and 2.5th--97.5th percentiles define the 68\% and 95\% pointwise envelopes, respectively.

\section{Matched-control construction and group-level comparison}
\label{app:matching_method}

We construct the control pools in the two-dimensional $\log_{10}(\mbh/M_{\odot})$--$\log_{10}\ledd$ space. Both quantities are standardised using the mean and standard deviation of the 1\,138-object S82 parent sample, and Euclidean distances are calculated between each CH quasar and every S82 parent quasar. The 50 nearest neighbours of each CH quasar define its control pool.

We generate $B=10\,000$ control pseudo-samples. For each pseudo-sample, one quasar is drawn with equal probability from each control pool. If the same quasar is selected more than once, the draw is discarded and repeated, so every accepted pseudo-sample contains 11 distinct controls. The same accepted pseudo-samples are used for all four DRW parameters and for the ensemble SF comparison.

For each DRW parameter, the comparison is performed in $\log_{10}$ space. Let $T_{\rm CH}$ denote the median of the 11 CH measurements and $T_b$ the median of the 11 controls in pseudo-sample $b$. The $B$ values of $T_b$ form the matched-control reference distribution. Let $T_{\rm ref}$ denote the median of this distribution. Its 16th--84th and 2.5th--97.5th percentiles define the central 68\% and 95\% reference intervals, respectively. We define the offset as the CH median minus the reference median,
\begin{equation}
\Delta T=T_{\rm CH}-T_{\rm ref}.
\end{equation}

The empirical two-sided $p$ value is calculated as
\begin{equation}
p=\min\left[
1,
2\min\left(
\frac{1+N(T_b\leq T_{\rm CH})}{B+1},
\frac{1+N(T_b\geq T_{\rm CH})}{B+1}
\right)
\right],
\end{equation}
where $N(\cdot)$ is the number of pseudo-samples satisfying the stated condition. The $+1$ correction accounts for the finite number of Monte Carlo realisations and prevents a zero empirical probability \citep{Phipson2010}. The empirical two-sided $p$ value quantifies how extreme the CH median is relative to the matched-control reference distribution.

\section{Control-pool balance diagnostics}\label{app:matching}

We evaluate the balance of the control pools in the two matching variables, $\log_{10}(\mbh/M_{\odot})$ and $\log_{10}\ledd$. We also examine $\log_{10}L_{\rm bol}$ as an additional check. The 550 control-source entries correspond to 433 unique S82 quasars. For these diagnostics, each S82 quasar is weighted by the number of control pools in which it appears. 

For a variable $x$, the standardised mean difference (SMD) is
\begin{equation}
\mathrm{SMD}(x)=
\frac{\overline{x}_{\rm control,w}-\overline{x}_{\rm CH}}
{\sqrt{\left(s_{\rm control,w}^{2}+s_{\rm CH}^{2}\right)/2}},
\end{equation}
where $\overline{x}_{\rm control,w}$ and $s_{\rm control,w}$ are the weighted mean and standard deviation of the controls, and $\overline{x}_{\rm CH}$ and $s_{\rm CH}$ are the corresponding quantities for the CH quasars. A positive SMD indicates a larger mean for the controls. The empirical cumulative distribution functions (ECDFs) provide an additional comparison of the distribution shapes.

\begin{figure*}[t]
\centering
\includegraphics[width=0.92\textwidth]{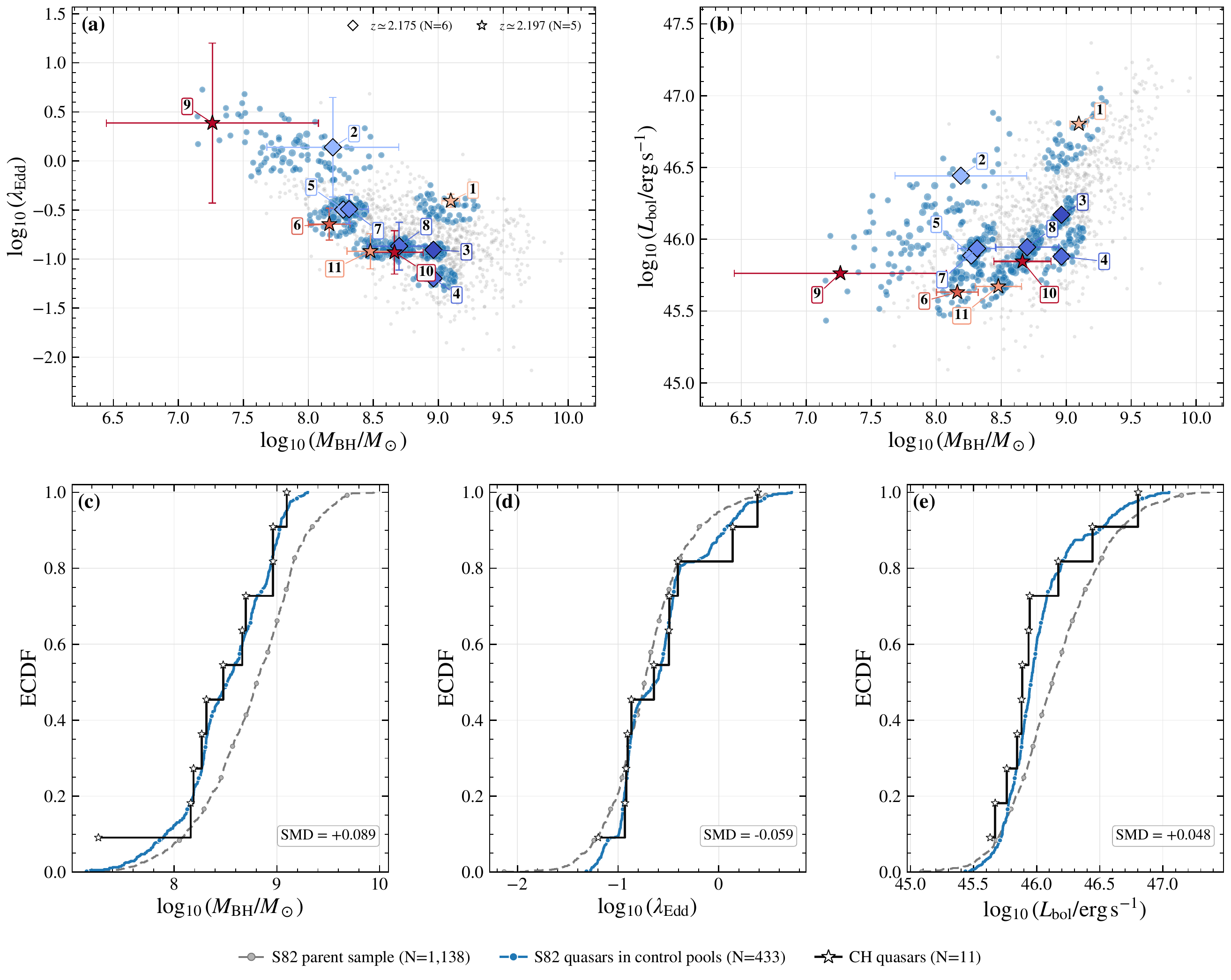}
\caption{Control-pool balance diagnostics. Panels~(a) and (b) show the 1\,138 S82 parent quasars (grey points), the 433 unique S82 quasars in the control pools (blue points), and the 11 CH quasars in the $\log_{10}(\mbh/M_{\odot})$--$\log_{10}\ledd$ and $\log_{10}(\mbh/M_{\odot})$--$\log_{10}L_{\rm bol}$ planes, respectively. Error bars show the catalogue uncertainties of the CH quasars. Panels~(c)--(e) show the ECDFs of the three quantities.}
\label{fig:matching}
\end{figure*}

As shown in Fig.~\ref{fig:matching}, the SMDs are $+0.089$, $-0.059$, and $+0.048$ for $\log_{10}(\mbh/M_{\odot})$, $\log_{10}\ledd$, and $\log_{10}L_{\rm bol}$, respectively. All three SMDs satisfy $|\mathrm{SMD}|<0.1$, indicating only small differences in the mean values \citep{Austin2009}, and the ECDFs show no clear differences in the overall distribution shapes. The control pools are therefore well balanced in the two matching variables, with $\log_{10}L_{\rm bol}$ showing similar consistency.

\section{Robustness to ZTF temporal sampling}\label{app:binning}

To further assess the effect of the different ZTF temporal sampling, we rebin the ZTF measurements in each S82 control light curve into fixed observer-frame bins of 35, 90, and 150~d. The SDSS and PS1 measurements remain unchanged. We then refit the $g$- and $r$-band light curves using the same DRW procedure and repeat the matched-control comparison.

\begin{figure*}[t]
\centering
\includegraphics[width=0.92\textwidth]{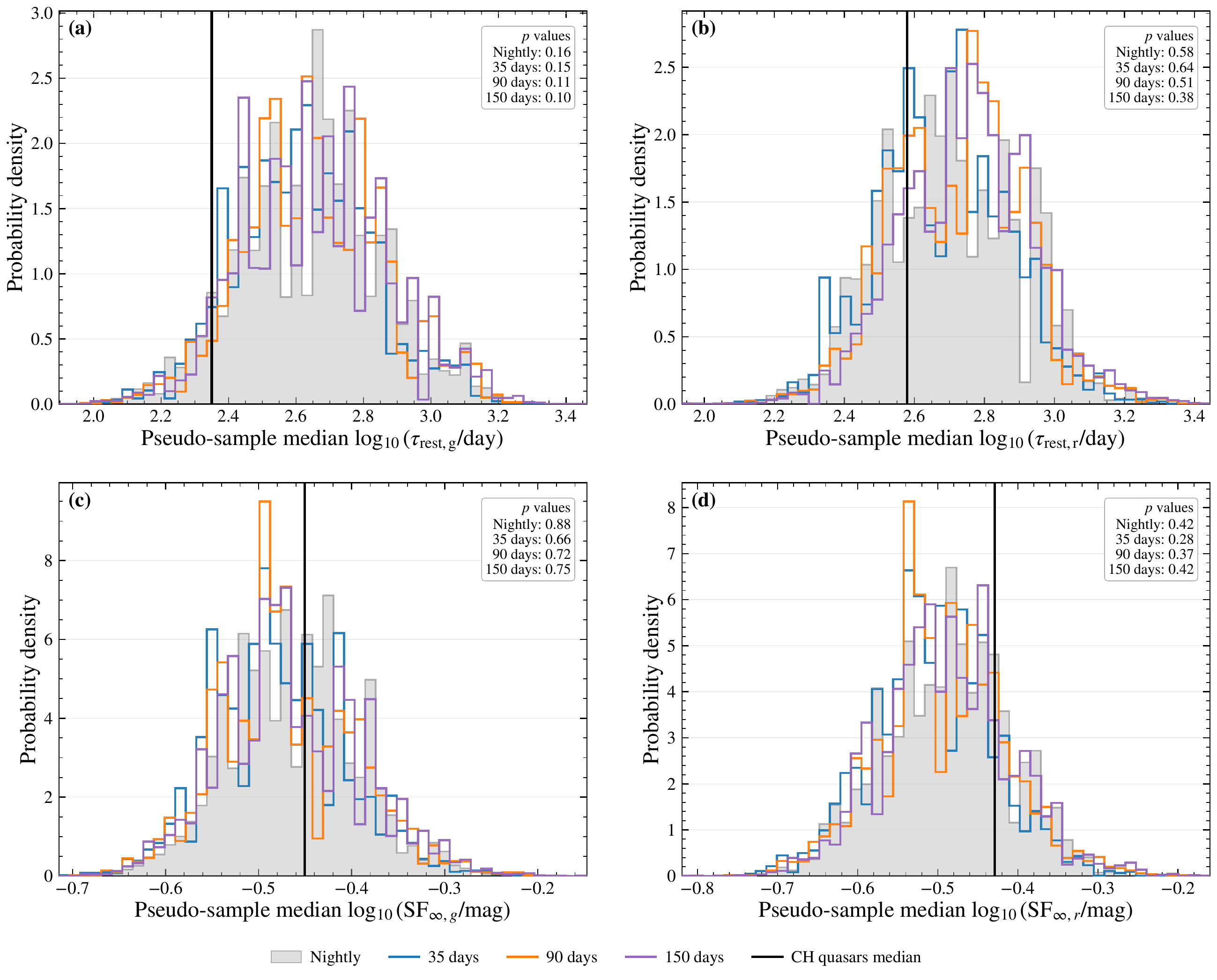}
\caption{Effect of rebinning the ZTF light curves of the S82 controls on the matched-control reference distributions. Panels~(a) and (b) show the distributions of the control pseudo-sample medians of $\log_{10}(\tau_{\rm rest}/{\rm d})$ in the $g$ and $r$ bands, and panels~(c) and (d) show the corresponding distributions of $\log_{10}(\sfinf/{\rm mag})$. Grey filled histograms show the original nightly analysis, and coloured step histograms show the analyses with 35, 90, and 150~d rebinning. Solid black lines mark the CH medians. The empirical two-sided $p$ values are listed in each panel.}
\label{fig:lcbin}
\end{figure*}

For the different bin sizes, the reference median $\tau_{\rm rest}$ values increase by approximately 5--9\%, and the reference median $\sfinf$ values change by less than 0.01~dex. All empirical two-sided $p$ values remain above 0.05. As shown in Fig.~\ref{fig:lcbin}, the rebinned reference distributions are similar to those from the original nightly analysis, and the population-level DRW conclusion is unchanged.

\end{document}